\documentclass[twocolumn,superscriptaddress,amsmath,amssymb,aps,pra]{revtex4}
\usepackage{amsmath,amsthm,amssymb}
\usepackage{latexsym}
\usepackage{amscd,graphicx}
\usepackage[titletoc]{appendix}
\usepackage{epsfig}
\usepackage{epstopdf}
\usepackage[normalem]{ulem}
\usepackage[dvipsnames]{xcolor}
\usepackage[colorlinks=true,linkcolor=red,citecolor=red]{hyperref}

\begin{document}

\newcommand{\cmt}[1]{{\textcolor{red}{#1}}}
\newcommand{\rvs}[1]{{\textcolor{blue}{#1}}}

\title{Strong Tunable Spin-Spin Interaction in a Weakly Coupled Nitrogen Vacancy Spin-Cavity Electromechanical System}

\author{Wei Xiong}

\affiliation{Department of Physics, Wenzhou University, Zhejiang 325035, China}
\affiliation{Interdisciplinary Center of Quantum Information and Zhejiang Province Key Laboratory of Quantum Technology and Device, Department of Physics and State Key Laboratory of Modern Optical Instrumentation, Zhejiang University, Hangzhou 310027, China}
\affiliation{School of Physics and Material Science, Anhui University, Hefei 230601, China}

\author{Jiaojiao Chen}
\affiliation{Hefei Preschool Education College, Hefei 230013, China}

\author{Baolong Fang}
\affiliation{Institute of Advanced Manufacturing Engineering, Hefei University, Hefei 230022, China}

\author{Mingfeng Wang}
\altaffiliation{mfwang@wzu.edu.cn}
\affiliation{Department of Physics, Wenzhou University, Zhejiang 325035, China}

\author{Liu Ye}
\altaffiliation{yeliu@ahu.edu.cn}
\affiliation{School of Physics and Material Science, Anhui University, Hefei 230601, China}

\author{J. Q. You}
\altaffiliation{jqyou@zju.edu.cn}
\affiliation{Interdisciplinary Center of Quantum Information and Zhejiang Province Key Laboratory of Quantum Technology and Device, Department of Physics and State Key Laboratory of Modern Optical Instrumentation, Zhejiang University, Hangzhou 310027, China}

\date{\today }

\begin{abstract}
	The long coherence time of a single nitrogen vacancy (NV) center spin in diamond is a crucial advantage for implementing quantum information processing. However, the realization of strong coupling between single NV spins is challenging. Here we propose a method to greatly enchance the interaction between two single NV spins in diamond which are only weakly coupled to an electromechanical cavity. Owing to the presence of a critical point for the linearized electromechanical subsystem, the coupling between a single NV spin and the high-frequency polariton (formed by the mechanical and cavity modes) can be fully decoupled, but the coupling between the single NV spin and the low-frequency polariton is however greatly enhanced. Thus, ac Stark shift of the single NV spin can be measured. With the low-frequency polariton as a quantum bus, a strong coupling between two single NV centers is achievable. This effective strong coupling can ensure coherent quantum-information exchange between two spin qubits in the weakly coupled spin-cavity elecromechanical system.
\end{abstract}


\maketitle

\section{Introduction}
A nitrogen vacancy (NV) center spin in diamond \cite{RS,doherty2013}, with long coherence time \cite{Gill2013, Jelezko2004,Balasubramian2009} and high tunability \cite{doherty2013}, is a promising qubit candidate for quantum information processing \cite{xiang2013,kurizki2015}.
However, spin-spin interaction is usually too weak to efficiently implement quantum-information exchange \cite{kubo2010,marcos2010,zhu2011}.
One popular method to overcome this drawback is the use of an ensemble containing a large number of NV spins \cite{Qiu-2014,Lu-2013,Mei-2009,Mei-2010}. Thus, the coupling strength between NV spin ensembles can be efficiently enhanced \cite{Raizen-89,Petersson-12,Rose-17}. However, it is difficult for the ensemble to implement direct single-qubit manipulation and the coherence time is also greatly shortened due to the inhomogeneous broadening \cite{Dobrovitski-08,Wesenberg-09,Julsgaard-13}. Another potential approach is to couple NV spins to the nanomechanical resonator \cite{Arcizet-2011,Bennett-2013,Chotorlishvili-2013,Li-2016,Rabl-2009,Rabl-2010,Kolkowitz-2012,Pigeau-2015,Li-2015,Zhou-2010,MacQuarrie-2013,Teissier-2014,Ovartchaiyapong-2014,Kepesidis-2013}, but the required strong magnetic gradient remains challenging experimentally \cite{Arcizet-2011,Kolkowitz-2012} and the strain force is inheretly tiny {for the ground states of NV spins~\cite{Teissier-2014,Ovartchaiyapong-2014}.}
Also, interacting spins with squeezed photons to enhance the coupling is proposed \cite{Lu-2015,Qin-2018,Xiong-2018}, but external noises may be introduced to the considered system.
{In addition, there are important efforts to couple remote NV spins via an optical network link~\cite{Humphreys-2018,Gao-2015} or a superconducting bus~\cite{Twamley2010}.}

Recently, optomechanical and electromechanical systems have attracted much interest in testing macroscopic quantum properties because of their appealing applications in quantum information science \cite{Kippenberg-2008,Marquardt-2009,Aspelmeyer-2012,Aspelmeyer-2014,Regal-2011,Teufel1-2011,Didier-2012,wangyindan,Vitali2007,clerk2010}. Based on remarkable progress in experiments \cite{Teufel2-2011,Chan-2011,Weis-2010,Safavi-Naeini-2011,Brooks-2012,Safavi-Naeini-2013,Bochmann-2013,Andrews-2014,Bagci-2014}, we can propose an experimentally accessible approach to realizing strong spin-spin coupling in a hybrid spin-cavity electromechanical system, where the single NV spin is only weakly coupled to the cavity mode. By applying a strong driving field to the cavity, two hybrid modes arising from the linearized strong coupling between the cavity and mechanical modes are generated, namely, the high-frequency and low-frequency polaritons. Tuning the linearized electromechanical coupling strength to a certain value (i.e., a critical point) by varying the driving field, we can have the electromechanical subsystem reach the {\it critical} regime. When operating the hybrid system around this critical point, the coupling between the single NV spin and the high-frequency polariton is totally suppressed if the cavity frequency detuning from the driving field is much larger than that of the mechanical resonator. However, the coupling strength between the NV spin and the low-frequency polariton is greatly {\it enhanced} more than three orders of magnitude of the single spin-cavity coupling. This strong coupling allows one to measure the ac Stark shift of a single NV spin. Taking the low-frequency polariton as a quantum bus, strong spin-spin coupling can be then induced. The results indicate that even for the weakly coupled spin-cavity electromechanical system, it is promising to probe the spin qubit states and implement polariton-mediated quantum information processing with single spin qubits.

\begin{figure}
	\center
	\includegraphics[scale=0.44]{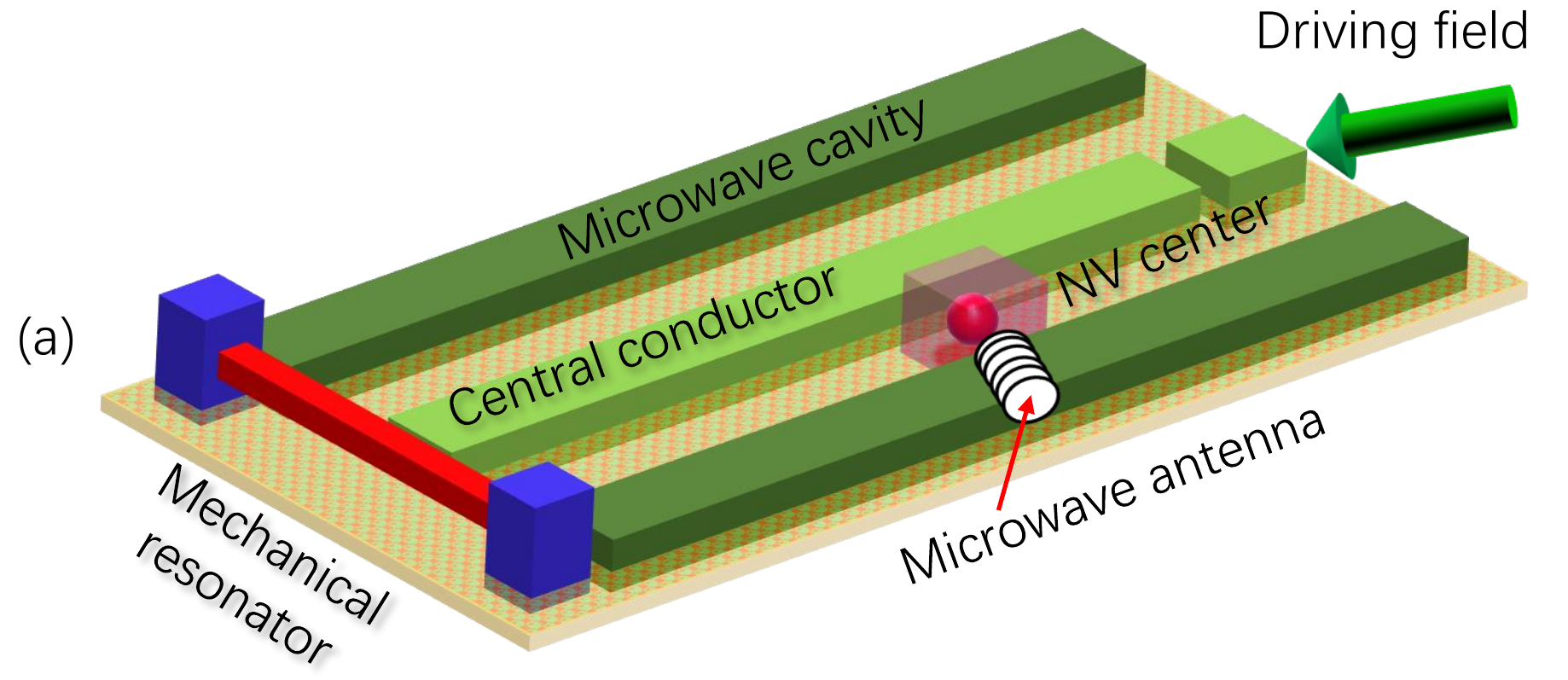}
	\includegraphics[scale=0.24]{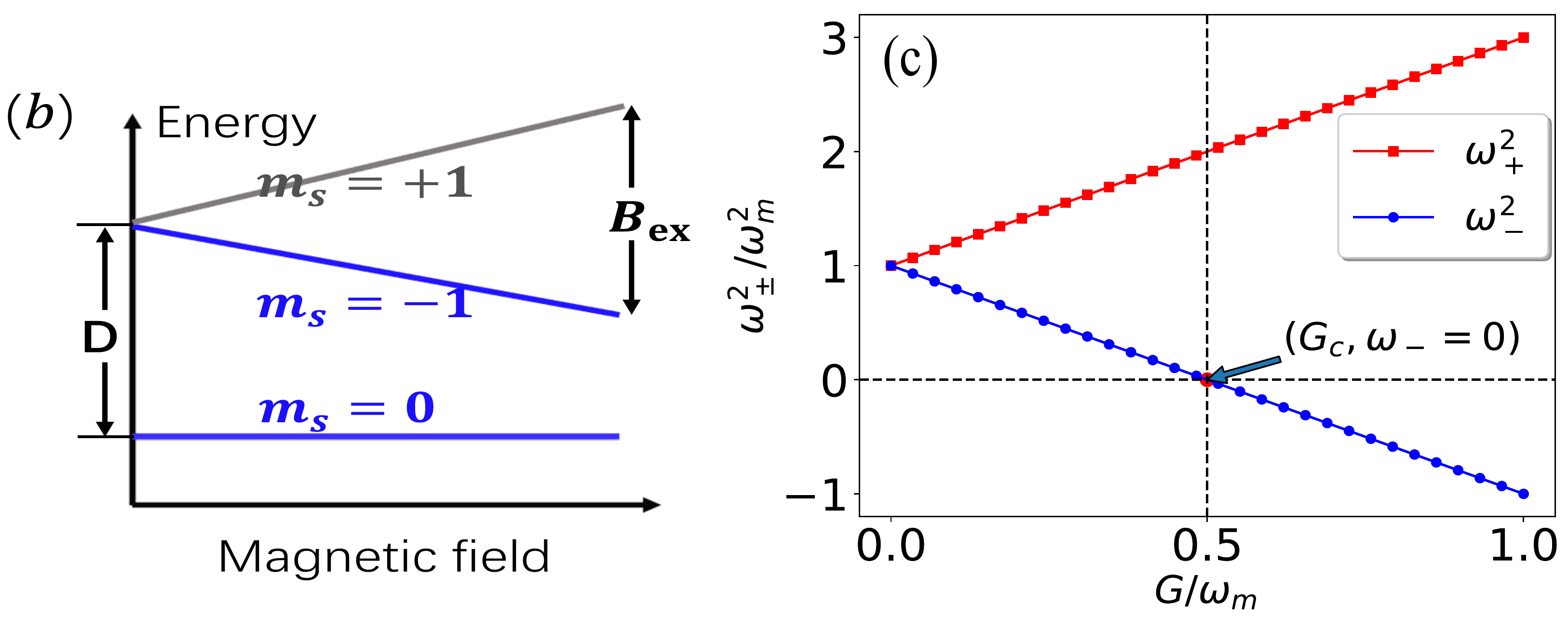}
	\caption{(a) Schematic diagram of the proposed hybrid system. The single NV center spin, located $d$ distance away from the central conductor, is weakly coupled to an electromechanical cavity and driven by a microwave field via a microwave antenna. This on-chip cavity is also driven by a strong microwave field. (b) The level structure of the triplet ground state $S=1$ in an NV center, where $D$ is the zero-field splitting and $B_{\rm ex}$ is an external magnetic field to lift the degenerate states $m_s=\pm1$. The lowest two sublevels $m_s=0,-1$ act as a two-level quantum system (i.e., qubit). (c) Frequencies of the two hybrid modes, generated by the strong coupling between the cavity and mechanical modes, versus the electromechanical coupling strength in units of the mechanical frequency.}
\label{fig1}
\end{figure}
\section{Model and Hamiltonian}
We consider a hybrid quantum system consisting of a {\it single} NV spin weakly coupled to an electromechanical cavity [Fig.\ref{fig1}(a)], where the NV spin in diamond with a spin $S=1$ triplet ground state [Fig.\ref{fig1}(b)] is located $d$ distance away from the central conductor of the coplanar-waveguide resonator.
Hereafter, we term this waveguide resonator as a (on-chip) cavity to distinguish it from the mechanical resonator. The Hamiltonian of the hybrid system can be written as (setting $\hbar=1$)
$H_{\rm tot}=H_{\rm NV}+H_{\rm EM}+H_{I}+H_D$,
where $H_{\rm NV}=\frac{1}{2}\omega_{\rm NV}\sigma_z$, with the transition frequency $\omega_{\rm NV}=D-g_e\mu_B B_{\rm ex}$ between the lowest two levels of the triplet ground state of the NV. Here, $D\approx2.87$ GHz is the zero-field splitting, $g_e=2$ is the Land\'{e} factor, $\mu_B$ is the Bohr magneton, and $B_{\rm ex}$ is the external magnetic field to lift the near-degenerate states $|m_s=\pm1\rangle$. The second term $H_{\rm EM}=\omega_a a^\dag a+\omega_m b^\dag b-ga^\dag a(b+b^\dag)$ is the Hamiltonian of the cavity electromechanical subsystem \cite{law1995}, where $\omega_a$ is the frequency of the cavity mode when the mechanical resonator is at its equilibrium position, $\omega_m$ is the frequency of the mechanical mode, and $g$ is the single-photon coupling strength between the cavity and mechanical modes.
The third term $H_I$ describes the magnetic coupling between the single NV spin and the cavity mode. Under the rotating-wave approximation,
$H_I=\lambda(a^\dag\sigma_-+a\sigma_+)$, with $\lambda=2 g_e \mu_B B_{0,{\rm rms}}(d)$ \cite{Twamley2010}, where $B_{0,{\rm rms}}(d)=\mu_0 I_{\rm rms}/2\pi d$, with $\mu_0$ being the permeability of vacuum and $I_{\rm rms}=\sqrt{\hbar\omega_a/2L_a}$.
To estimate $\lambda$, $\omega_a\sim 2\pi\times 2$ GHz and $L_a\sim 2$ nH are chosen \cite{Niemczyk2009}. For $d\sim50$~$\mu$m, $\lambda\sim 2\pi\times 70$~Hz, and $d\sim50$~nm gives $\lambda\sim 2\pi\times 7$~KHz. Obviously, the estimated spin-cavity coupling strength is smaller than the typical decay rate of the cavity with the gigahertz frequency and quality factor $Q\sim 3\times 10^4$ \cite{Niemczyk2009,Niemczyk2010}, i.e., $\lambda<\kappa=\omega_a/Q\sim 1$ MHz. This indicates that the spin-cavity coupling is in the {\it weak-coupling} regime. The last term, $H_D=\Omega_d a^\dag\exp(-i\omega_d t)+\Omega_d^*a \exp(i\omega_d t)-\Omega_{\rm NV}\sigma_+\exp(-i\omega_d t)-\Omega_{\rm NV}^*\sigma_- \exp(i\omega_d t)$, involves the driving fields acting on the cavity mode and the NV spin. Due to the large $\Omega_d$, the electromechanical coupling strength can be amplified to $G\equiv g\sqrt{N}>\kappa$, where $N=|\langle a\rangle|^2$ is the intracavity mean photon number. This amplification is achieved by linearizing the Hamiltonian with $a=\langle a\rangle+\delta a$ and $b=\langle b\rangle+\delta b$, where $\langle a\rangle=-\Omega_d/(\Delta_a-i\kappa)$ and $\langle b\rangle=g |\langle a\rangle|^2/(\omega_m-i\gamma_m)$ are the mean values of the operators $a$ and $b$ in the steady state, respectively~\cite{sm}, with $\Delta_a=\omega_a-\omega_d-g(\langle b\rangle+\langle b\rangle^*)$ and $\gamma_m$ being the decay rate of the mechanical mode. As the driving field on the cavity will indirectly cause the NV spin to flip, an additional microwave field with frequency $\omega_d$ and amplitude $\Omega_{\rm NV}$ is imposed to the NV spin via a microwave antenna [see Fig.\ref{fig1}(a)], so as to cancel this flip. Then, the linearized Hamiltonian of the hybrid spin-cavity elecromechanical system is given by~\cite{sm}
{
 \begin{align}
H_{\rm lin}=&\frac{1}{2}\Delta_{\rm NV}\sigma_z+H_{\rm LEM}+\lambda(\delta a^\dag\sigma_-+\delta a\sigma_+),
\label{eq1}
 \end{align}
where $\Delta_{\rm NV}=\omega_{\rm NV}-\omega_d$, and $H_{\rm LEM}=\Delta_a \delta a^\dag \delta a+\omega_m \delta b^\dag \delta b-G(\delta a+\delta a^\dag)(\delta b+\delta b^\dag)$ is the linearized Hamiltonian of the cavity electromechanical subsystem.
In deriving Eq.~(\ref{eq1}), we assume $\lambda \langle a\rangle=\Omega_{\rm NV}$~\cite{sm}, where $\langle a\rangle$ is controlled by $\Omega_d$ and can be determined via the input-output theory~\cite{walls} by probing the output field of the cavity. There were proposals for measuring a weak coupling strength $\lambda$~\cite{zengzhao,bin2018,carreno2015} and they were demonstrated experimentally~\cite{Tekavec2007,cuevas}. With a given $\lambda$, the condition $\lambda \langle a\rangle=\Omega_{\rm NV}$ can be readily satisfied, because both $\Omega_d$ and $\Omega_{\rm NV}$ are precisely tunable.
Here, the effects of the driving fields are included in $\Delta_a$ and $G$, while quantum behaviors of the system are kept in the linearized Hamiltonian~(\ref{eq1}) via the fluctuation operators $\delta a$ and $\delta b$.
In the linearized Hamiltonian $H_{\rm LEM}$, there are counter-rotating terms $G\delta a\delta b$ and $G\delta a^\dag\delta b^\dag$, which are related to the squeezing effect in the system. As shown below, when $G$ approaches its critical value, these counter-rotating terms play an increasingly important role~\cite{Sudhir2012} and polaritons of novel properties can be formed.}

\section{Strong coupling between a single NV spin and the low-frequency polariton}
Assisted by a strong driving field, the mechanical mode can strongly couple to the cavity mode via the linearized electromechanical coupling strength $G$, yielding two hybrid modes (i.e., polaritons) with eigenfrequencies
 \begin{align}
 \omega _ \pm ^2 =\frac{1}{2} \Big[\Delta _a^2 + \omega _m^2 \pm \sqrt {{{(\Delta _a^2 - \omega _m^2)}^2} + 16{G^2}{\Delta _a}{\omega _m}}\Big] ,\label{eq2}	
 \end{align}
where $\omega_+$ and $\omega_-$ correspond to the high- and low-frequency polaritons, respectively. These eigenfrequencies are directly obtained by diagonalizing the linearized Hamiltonian $H_{\rm LEM}$ in Eq.~(\ref{eq1}) (see \cite{sm} for details). With the eigenvectors of the hybrid modes, Eq. (\ref{eq1}) becomes
\begin{align}
\mathcal{H}=&\frac{1}{2}\Delta_{\rm NV}\sigma_z+\omega_+a_+^\dag a_++\omega_-a_-^\dag a_-\notag\\
&+\lambda_+(a_-^\dag\sigma_-+a_-\sigma_+)+\lambda_-(a_-^\dag\sigma_++a_-\sigma_-)\notag\\
&+\eta_+(a_+^\dag\sigma_-+a_+\sigma_+)+\eta_-(a_+^\dag\sigma_++a_+\sigma_-),\label{eq3}
\end{align}
where $\lambda_\pm=\lambda\cos\theta(\Delta_a\pm\omega_-)/2\sqrt{\Delta_a\omega_-}$ denotes the effective coupling strength between the NV spin and the low-frequency polariton, and $\eta_\pm=\lambda\sin\theta(\Delta_a\pm\omega_+)/2\sqrt{\Delta_a\omega_+}$ is the effective coupling strength between the NV spin and the high-frequency polariton. The parameter $\theta$ is defined by $\tan(2\theta)=4G\sqrt{\Delta_a\omega_m}/(\Delta_a^2-\omega_m^2)$.
Both $\lambda_\pm$ and $\eta_\pm$ can be {\it tuned} by the driving field on the cavity.
From Eq. (\ref{eq2}), one can see that $\omega_+^2$ increases with the linearized optomechanical coupling $G$, but $\omega_-^2$ decreases [see Fig.~\ref{fig1}(c)]. In particular, when $G$ reaches a certain value,
\begin{equation}
G=G_c\equiv \frac{1}{2}\sqrt{\Delta_a\omega_m},
\end{equation}
critical phenomenon occurs~\cite{lu2013sr}, where the frequency of the low-frequency polariton vanishes (i.e., $\omega_-=0$) [see the red point in Fig.~\ref{fig1}(c)]. {Here we consider the case with both $G\rightarrow G_c$ (i.e., $\omega_-\rightarrow 0$) and $\Delta_a/\omega_m\gg1$, where $\lambda_+\approx\lambda_-\rightarrow\frac{\lambda}{2}\sqrt{\Delta_a/\omega_-}\gg\lambda,~\eta_+\rightarrow \lambda\omega_m/\Delta_a\ll \lambda$, and $\eta_-\rightarrow 0$. This indicates that the coupling between the NV spin and the high-frequency polariton can be ignored in Eq.~(\ref{eq3}). Correspondingly,  the coupling between the {\it single} NV spin and the low-frequency polariton is greatly {\it enhanced} due to the extremely small value of $\omega_-$ and tunable parameter $\Delta_a$.
Thus, raising the photon occupation number in the cavity to have $G\equiv g\sqrt{N}$ reach $G_c$, we can use the critical behavior of the effective spin-polariton system to enhance the coupling between the spin and the low-frequency polariton.}
\begin{figure}
	\includegraphics[scale=0.28]{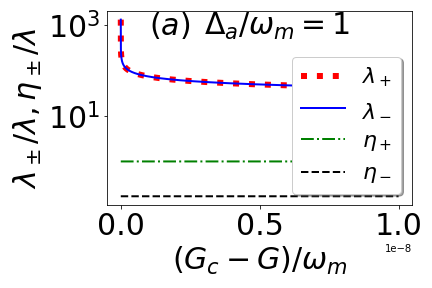}
	\includegraphics[scale=0.28]{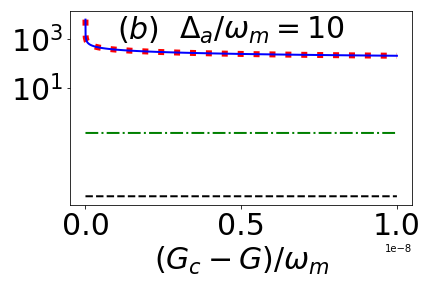}
	\includegraphics[scale=0.28]{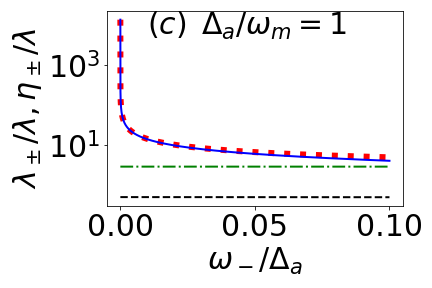}
	\includegraphics[scale=0.280]{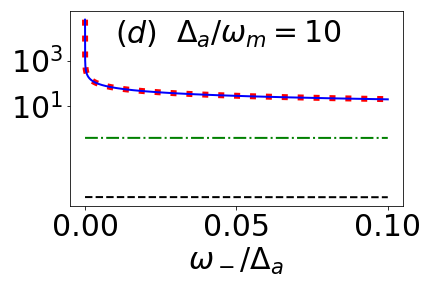}
	\caption{Coupling strength between the NV spin and the high- (low-) frequency polaritons versus the dimensionless parameters $(G_c-G)/\omega_m$ and $\omega_-/\Delta_a$.  Here $\Delta_a/\omega_m=1$ in both (a) and (c), and $\Delta_a/\omega_m=10$ in both (b) and (d). }\label{fig2}
\end{figure}

In Fig.~\ref{fig2}, we plot the coupling strength between the NV spin and the high- (low-) frquency polariton versus the dimensionless parameters $(G_c-G)/\omega_m$ and $\omega_-/\Delta_a$ with $\Delta_a/\omega_m=1,~10$. When $G\rightarrow G_c$ or $\omega_-\rightarrow 0$, the coupling between the NV spin and the low-frequency polariton can be extremely strong [see Figs.~\ref{fig2}(a) and \ref{fig2}(c)]. With increasing $\Delta_a$, $\lambda_\pm$ can be further enhanced [see Figs.~\ref{fig2}(b) and \ref{fig2}(d)].  In principle, $\omega_-$ can be very close to zero and the detuning $\Delta_a$ can be sufficiently large. These can yield $\lambda_\pm$ in the {\it strong-coupling} regime (i.e., $\lambda_\pm\geq\kappa$) to allow coherent quantum-information exchange between the single NV spin and the low-frequency polariton. In this regime, rotating-wave approximation is still valid and the counter-rotating term related to $\lambda_-$ in Eq. (\ref{eq3}) can be safely ignored. Thus, around the critical point $G=G_c$ and when $\Delta_a\gg\omega_m$, the Hamiltonian in Eq. (\ref{eq1}) reduces to the Janes-Cumming model,
\begin{align}
\mathcal{H}_{\rm JC}=&\frac{1}{2}\Delta_{\rm NV}\sigma_z+\omega_-a_-^\dag a_-
+\lambda_+(a_-^\dag\sigma_-+a_-\sigma_+),\label{eq4}
\end{align}
when considering the interaction between the single NV spin and the low-frequency polariton. To estimate $\lambda_+$, we choose $\Delta_a=10^6\omega_-$, then $\lambda_+/\lambda=0.5\times10^3$, which indicates that $\lambda_+$ can be approximately enhanced three orders of magnitude of $\lambda$ in our hybrid system. Specifically, with $\lambda=2\pi\times7$ KHz, as estimated above for $d=50$~nm, $\lambda_+\sim 2\pi\times 3.5$ MHz, {larger than} the decay rate of the gigahertz cavity with quality factor~$\sim 10^4$. This implies that the coupling between the single NV spin and the low-frequency polariton is in the strong-coupling regime. Theoretically, the value of $\lambda_+$ can be even larger due to the extremely small $\omega_-$.

\begin{figure}
	\includegraphics[scale=0.4]{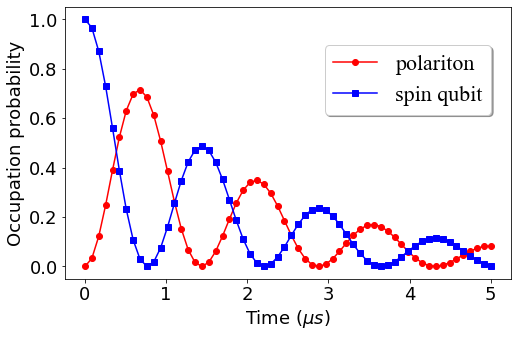}
	\caption{The mean occupation number of the low-frequency polariton and the occupation probability of the NV spin versus the evolution time, where the polariton is initially prepared in the ground state and the spin qubit is in the excited state. We choose $\lambda_+=2\pi\times 3.5$ MHz, $\gamma_\perp=1$ KHz, and $\kappa=1$ MHz.}
\label{fig3}
\end{figure}

When decoherence is considered, the dynamics of the above low-frequency polariton-spin system can be governed by a master equation
\begin{equation}
\frac{d\rho}{dt}=-i[\mathcal{H}_{\rm JC},\rho]+\kappa \mathcal{D}[a_-]\rho
+\gamma_\perp\mathcal{D}[\sigma_-]\rho+\gamma_\parallel\mathcal{D}[\sigma_z]\rho,\label{eq7}
\end{equation}
where $\mathcal{D}[o]\rho=o\rho o^\dag-\frac{1}{2}(o^\dag o\rho+\rho o^\dag o)$ for a given operator $o$, and $\gamma_\perp$ ($\gamma_\parallel$) is the transversal (longitudinal) relaxation rate of the NV spin. Experimentally, $\gamma_\perp\gg\gamma_\parallel$ \cite{Angerer17}, so the longitudinal relaxation rate can be ignored. To solve Eq.~(\ref{eq7}), we choose $\gamma_\perp=1$ KHz and $\kappa=1$ MHz for example. By individually tuning the driving fields acting on the cavity and the NV spin, we can have $\Delta_{\rm NV}=\omega_-$ in Eq.~(\ref{eq4}). In Fig.~\ref{fig3}, we show the time evolutions of the mean occupation number $\langle a^\dag_-a_-\rangle$ of the low-frequency polariton and the occupation probability of the NV spin qubit. Initially, the low-frequency polariton is prepared in the ground state and the NV spin is in the excited state $|m_s=-1\rangle\equiv|1\rangle$. It can be seen that Rabi oscillations between the low-frequency polariton and the NV spin occur owing to the strong coupling, in spite of the decoherence in the hybrid system.

\section{Effective strong coupling between two NV spins and the ac Stark shift}
We consider two separated NV center spins coupled to the low-frequency polariton mode with coupling strengths $\lambda_{+}^{(1)}$ and $\lambda_{+}^{(2)}$. We can use this low-frequency polariton as a quantum interface to achieve an effective strong spin-spin coupling in the dispersive regime, i.e., $\zeta_i=\lambda_{+}^{(i)}/|\delta_i|\ll 1$, where $\delta_i=\Delta_{\rm NV}^{(i)}-\omega_-$, $i=1,2$, are the frequency detunings of the two NV spins from $\omega_-$. With the Fr\"{o}hlich-Nakajima transformation \cite{Frohlich,Nakajima}, the effective spin-spin Hamiltonian can be obtained as~\cite{sm}
\begin{align}
\mathcal{H}_{\rm eff}=\sum_{i=1}^2\frac{1}{2}\Delta_{\rm eff}^{(i)}\sigma_z^{(i)}+g_{\rm eff}(\sigma_+^{(1)}\sigma_-^{(2)}+\sigma_-^{(1)}\sigma_+^{(2)}),\label{eq6}
\end{align}
where $\Delta_{\rm eff}^{(i)}=\Delta_{\rm NV}^{(i)}+\lambda_+^{(i)}\zeta_i(1+2N_{\rm pl})$ are the effective transition frequencies of the two NV spins, which depend on the mean occupation number $N_{\rm pl}=\langle a_-^\dag a_-\rangle$ of the low-frequency polariton, and $g_{\rm eff}=\frac{1}{2}(\lambda_+^{(1)}\zeta_2+\lambda_+^{(2)}\zeta_1)$ is the effective spin-spin coupling strength induced by the low-frequency polariton.
This coupling can allow coherent quantum-information exchange between the two separated NV spins
when they are in the strong-coupling regime.
For convenience and without loss of generality, we assume $\lambda_{+}^{(1)}=\lambda_{+}^{(2)}=\lambda_+=2\pi\times3.5$ MHz for $d=50$ nm and $\delta_1=\delta_2=2\pi\times35$ MHz, which leads to $g_{\rm eff}=2\pi\times 350$ KHz~$\sim 2.2$ MHz. This value is comparable to the decay rate of the gigahertz microwave cavity with quality factor $Q=10^4$. In fact, a microwave cavity with higher quality factor (e.g., $Q=10^5$) can be fabricated experimentally \cite{goppl2008}. Therefore, it is reasonable to conclude that the effective spin-spin coupling can reach the strong-coupling regime. {To obtain this strong coupling, we assume  $d=50$~nm. With current nanofabrication technology, it has been achieved to place an NV spin within $\sim30$ nm inside an on-chip cavity of length $\sim 1$~cm~\cite{siampour}. Therefore, our proposal is experimentally feasible. Owing to the larger size of the cavity compared to a single NV spin, it allows one to place multiple spins in a cavity for a scalable network.}

Furthermore, we show that the weakly coupled spin-cavity electromechanical system allows one to probe the ac Stark shift of the spin energy level induced by a single excitation of the low-frequency polariton, i.e., $N_{\rm pl}=\langle a^\dag_-a_-\rangle=1$. To obtain this, we consider that the spin is {\it dispersively} coupled to the low-frequency polariton, where $\zeta=\lambda_+/|\delta|\ll1$, with $\delta=\Delta_{\rm NV}-\omega_-$. Under this condition, the Fr\"{o}hlich-Nakajima transformation can be used to diagonalize the Hamiltonian (\ref{eq4}) to
\begin{align}
\mathcal{H}_{\rm AC}=\frac{1}{2}(\Delta_{\rm NV}+\lambda_+\zeta+2\lambda_+\zeta a_-^\dag a_-)\sigma_z+\omega_-a_-^\dag a_-,
\end{align}
where $\zeta=\lambda_+/|\delta|$ is the zero-point energy of the low-frequency polariton. Obviously, the frequency of the NV spin is shifted by both the zero-point energy $\lambda_+\zeta$ and the low-frequency polariton-dependent Stark shift $2\lambda_+\zeta N_{\rm pl}$. At the single polariton excitation $N_{\rm pl}=1$, the Stark shift is $2\lambda_+\zeta$. To estimate this Stark shift, {we choose $\lambda_+=2\pi\times 3.5$ MHz, $\delta=2\pi\times35$ MHz  and $\kappa=1$ MHz. These parameters ensure that the dispersive condition is valid and gives rise to  $2\lambda_+\zeta\sim2\pi\times0.7$ MHz}. As shown above, the value of  $\lambda_+$ can be much larger with increasing $\Delta_a/\omega_-$, so the Stark shift induced by a single polariton excitation is observable. Comparing with strong-coupling cases \cite{wang-2020,forn-2010}, our proposal grealty reduces experimental difficulties. It provides a promising way to realize coherent quantum-information exchange between two separated NV spins and can be used to probe the ac Stark shift of the single NV spin in the weakly coupled spin-cavity electromechanical systems.

\begin{figure*}
	\includegraphics[scale=0.4]{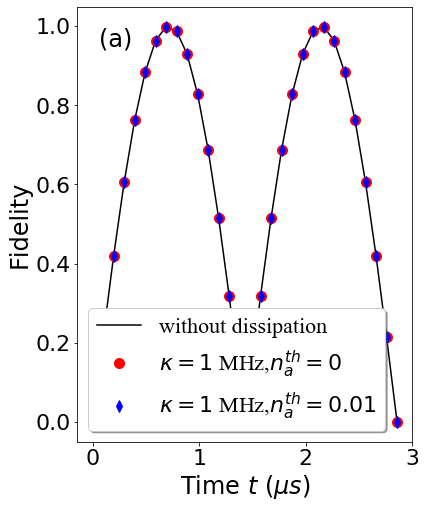}
	\includegraphics[scale=0.4]{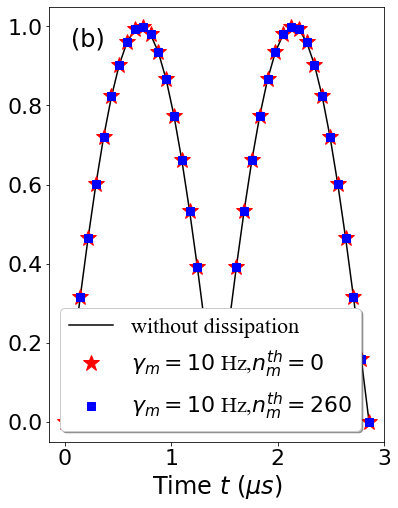}
	\includegraphics[scale=0.4]{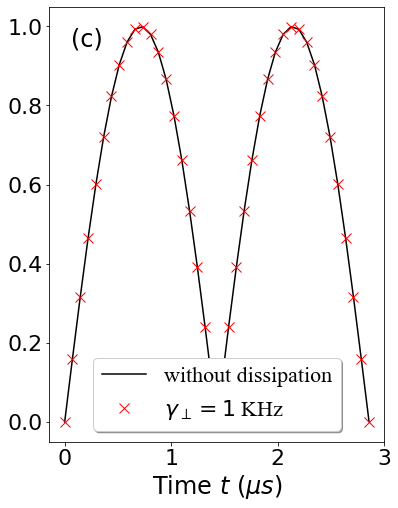}
	\caption{(Color online) The fidelity of the iSWAP gate versus the evolution time without and with (a) the cavity's decay rate at zero ($n_c^{\rm th}=0$) and nonzero ($n_c^{\rm th}=0.01$, i.e., $T\sim 20$~mK) temperatures, (b) the decay rate of the mechancial resonator at zero ($n_m^{\rm th}=0$) and nonzero ($n_m^{\rm th}=260$, i.e., $T\sim 20$~mK) temperatures, and (c) the transversal relaxation rate. Here we choose the parameters $\omega_-=100$ Hz, $\Delta_a=10^6 \omega_-$, $\omega_m=10^5 \omega_-$, $\lambda=2\pi\times7$ KHz, $\lambda_+^{(1)}=\lambda_+^{(2)}=\frac{1}{2}\lambda\sqrt{\Delta_a/\omega_-}$, $\Delta_{\rm NV}^{(1)}=\Delta_{\rm NV}^{(2)}=\delta=10\lambda_+^{(1)}$, and $g_{\rm eff}=[\lambda_+^{(1)}]^2/\delta$.}\label{fig}
\end{figure*}

{With the effective Hamiltonian (\ref{eq6}) [i.e., Eq.~(7) in the main text], the two-qubit iSWAP gate can be realized at $\Delta_{\rm eff}^{(1)}\approx\Delta_{\rm eff}^{(2)}$. In the interaction picture, the Hamiltonian (\ref{eq6}) becomes
\begin{align}\label{s59}
	\mathcal{H}_{\rm eff}^I=g_{\rm eff}(\sigma_-^{(1)}\sigma_+^{(2)}+\sigma_+^{(1)}\sigma_-^{(2)}).
\end{align}
The corresponding time evolution operator reads
\begin{align}
	\mathcal{U}(t)=\exp(-i\mathcal{H}_{\rm eff}^I t).
\end{align}
The final state of the two-spin system described by Eq.~(\ref{s59}) is
\begin{align}
	|\Psi_t\rangle=\mathcal{U}(t)|\Psi_0\rangle,
\end{align}
where $|\Psi_0\rangle$ is the initial state of the two-spin system. Specifically,
\begin{align}\label{s62}
	|\Psi_0\rangle&=|g_1\rangle|g_2\rangle \rightarrow |\Psi_t\rangle=|g_1\rangle|g_2\rangle,\nonumber\\
	|\Psi_0\rangle&=|g_1\rangle|e_2\rangle \rightarrow\nonumber\\
	|\Psi_t\rangle &=\cos(g_{\rm eff}t)|g_1\rangle|e_2\rangle-i\sin(g_{\rm eff}t)|e_1\rangle|g_2\rangle,\nonumber\\
	|\Psi_0\rangle&=|e_1\rangle|g_2\rangle \rightarrow\nonumber\\
	|\Psi_t\rangle&=\cos(g_{\rm eff}t)|e_1\rangle|g_2\rangle-i\sin(g_{\rm eff}t)|g_1\rangle|e_2\rangle,\nonumber\\
	|\Psi_0\rangle&=|e_1\rangle|e_2\rangle \rightarrow |\Psi_t\rangle=|e_1\rangle|e_2\rangle.
\end{align}
When $g_{\rm eff}t=\pi/2$, Eq.~(\ref{s62}) reduces to
\begin{align}
	|g_1\rangle|g_2\rangle&\rightarrow|g_1\rangle|g_2\rangle,\notag\\
	|g_1\rangle|e_2\rangle&\rightarrow -i|e_1\rangle|g_2\rangle,\notag\\
	|e_1\rangle|g_2\rangle&\rightarrow -i|g_1\rangle|e_2\rangle,\notag\\
	|e_1\rangle|e_2\rangle&\rightarrow |e_1\rangle|e_2\rangle,\label{s66}
\end{align}
which is just the iSWAP gate.

For a realistic gate, the dissipations in the on-chip cavity, mechanical resonator and two spins should be included.
Near the critical point (i.e., $G\rightarrow G_c$), $\omega_-\rightarrow 0$. Hence, the annihilation operator of the low-frequency polariton~\cite{sm} reduces to
\begin{align}\label{s65}
	a_-=\frac{1}{2}\bigg[\cos\theta\sqrt{\frac{\Delta_a}{\omega_-}}(\delta a-\delta a^\dag)-\sin\theta\sqrt{\frac{\omega_m}{\omega_-}}(\delta b-\delta b^\dag)\bigg].
\end{align}
By substituting Eq.~(\ref{s65}) into Eq.~(\ref{eq6}), the quantum dynamics of the considered system can be given by
\begin{align}
	\frac{d\rho}{dt}&=-i[\mathcal{H}_{\rm eff},\rho]+\kappa (n^{\rm th}_a+1) \mathcal{D}[\delta a]\rho+\kappa n^{\rm th}_a \mathcal{D}[\delta a^\dag]\rho\nonumber\\
	+&\gamma_m (n^{\rm th}_m+1) \mathcal{D}[\delta b]\rho+\gamma_m n^{\rm th}_m \mathcal{D}[\delta b^\dag]\rho+\gamma_\perp\mathcal{D}[\sigma_-]\rho,\label{s67}
\end{align}
{where $\mathcal{H_{\rm eff}}$ is given by Eq.~(\ref{eq6}) and $n_{a(m)}^{\rm th}=[\exp(\hbar\omega_{a(m)}/K_B T)-1]^{-1}$ is the thermal photon (phonon) occupation in the cavity (mechanical) mode}.

Figure~\ref{fig} show the numerical simulation for the fidelity of the iSWAP gate obtained using the effective Hamiltonian in Eq.~(\ref{eq6}) without and with dissipations in the cavity, mechanical resonators and NV spins. Here we choose $\omega_-=100$ Hz, $\Delta_a=10^6 \omega_-$, $\omega_m=10^5 \omega_-$, $\lambda=2\pi\times7$ KHz, $\lambda_+^{(1)}=\lambda_+^{(2)}=\frac{1}{2}\lambda\sqrt{\Delta_a/\omega_-}$, $\Delta_{\rm NV}^{(1)}=\Delta_{\rm NV}^{(2)}=\delta=10\lambda_+^{(1)}$, and $g_{\rm eff}=[\lambda_+^{(1)}]^2/\delta$. These choices of the parameters ensure the harnessed approximation valid. As the initial state, we prepare one spin qubit in its ground state, another spin qubit in the excited state, and both the cavity and the mechanical resonator are in the thermal state. In Fig.~\ref{fig}(a), we plot the iSWAP-gate fidelity versus the evolution time for different cavity decay rates (i.e., $\kappa=0,~1$ MHz) at zero ($n^{\rm th}_a=0$) and nonzero ($n^{\rm th}_a=0.01$, i.e., $T\sim 20$ mK) temperatures, respectively. Compared with the results at zero temperature, it can bee seen that the fidelity of the iSWAP gate is hardly affected by cavity's decay rate at $T\sim 20$~mK. Actually, this temperature is close to the temperatures often used for experiments on the on-chip cavity (e.g., the coplanar waveguide resonator). In Fig.~\ref{fig}(b), we plot the iSWAP-gate fidelity versus the evolution time for different decay rates of the mechanical resonator (i.e., $\gamma_m=0,~10$ Hz) at zero ($n^{\rm th}_m=0$) and nonzero ($n^{\rm th}_m=260$, i.e., $T\sim 20$~mK) temperatures. It is shown that the iSWAP-gate fidelity is strongly robust against the dissipation in the mechanical resonator. Also, we plot the iSWAP-gate fidelity versus the evolution time by considering the transversal relaxation of the NV spins. For simplicity, we assume the two NV spins have the same transversal relaxation, i.e., $\gamma_\perp=1$ KHz. Since $\gamma_\perp\ll g_{\rm eff}$, the gate fidelity is almost not affected by it. In short, Fig.~\ref{fig} shows that the iSWAP-gate fidelity is robust against the dissipations in the cavity, mechanical resonator and spins. This is due to the fact that the polariton formed by the cavity and mechanical modes is dispersively coupled to the spins and a strong coupling between the two separate spins can be induced in comparison with the decay rates of the cavity and spins.}

\section{Conclusion}
We have proposed a method to realize a strong tunable spin-spin coupling in a hybrid quantum system consisting of NV spins weakly coupled to an elecromechanical cavity. By taking advantage of the critical behavior of the linearized electromechanical system, the high-frequency polariton can be {\it decoupled} from the NV spin, but the coupling between the low-frequency polariton and the NV spin can be greatly {\it enhanced}. With experimentally accessible parameters, this coupling can reach the strong-coupling regime. Thus, ac Stark shift of the single NV spin induced by the single excitation of the low-frequency polariton can be resolved and the polariton-mediated quantum-information exchange between spins can be realized. Our proposal can provide a feasible way to probe spin qubit states and implement polariton-mediated quantum information processing with single spin qubits in the weakly coupled spin-cavity electromechanical systems.

\begin{acknowledgments}
This work is supported by the National Key Research and Development Program of China (Grant No.~2016YFA0301200), the National Natural Science Foundation of China (Grants No.~U1801661, No.~11934010, and No.~11804074),
Talent Development Funding of Hefei University (Grant No.~18-19RC60), Candidates of Hefei University Academic Leader (Grant No.~2016dtr02), and the Major Project of Hefei Preschool Education College (Grant No.~hyzzd2018003).
\end{acknowledgments}


\begin{thebibliography}{99}
	\bibitem{RS}R. Schirhagl, K. Chang, M. Loretz, and C. L. Degen, Nitrogen-Vacancy Centers in Diamond: Nanoscale Sensors for Physics and Biology, Annu. Rev. Phys. Chem. {\bf 65}, 83 (2014).
	
	\bibitem{doherty2013}M. W. Doherty, N. B. Manson, P. Delaney, F. Jelezko, J. Wrachtrupe, L. C. L. Hollenberg, The nitrogen-vacancy colour centre in diamond, Phys. Rep. {\bf 528}, 1 (2013).
	
	\bibitem{Gill2013}N. Bar-Gill, L. Pham, A. Jarmola, D. Budker, and R. Walsworth, Solid-state electronic spin coherence time approaching one second. Nat. Commun. {\bf4}, 1743 (2013).
	
	\bibitem{Jelezko2004}F. Jelezko, T. Gaebel, I. Popa, A. Gruber, and J. Wrachtrup, Observation of Coherent Oscillations in A Single Electron Spin, Phys. Rev. Lett. {\bf92}, 076401 (2004).
	
	\bibitem{Balasubramian2009}G. Balasubramian, P. Neumann, D. Twitchen, M. Markham, R. Koselov, N. Mizuochi, J. Isoya, J. Achard, J. Beck, J. Tissler, V. Jacques, P. R. Hemmer, F. Jelezko, and J. Wrachtrup, Ultralong spin coherence time in isotopically engineered diamond, Nat. Mater. {\bf8}, 383 (2009).
	
	\bibitem{xiang2013}Z. L. Xiang, S. Ashhab, J. Q. You, and F. Nori, Hybrid quantum circuits: Superconducting circuits interacting withother quantum systems, Rev. Mod. Phys.{\bf85}, 623 (2013).
	
	\bibitem{kurizki2015}G. Kurizki, P. Bertet, Y. Kubo, K. M\o{}lmer, D. Petrosyan, P. Rabl, and J. Schmiedmayer, Quantum technologies with hybrid systems, Proc. Natl. Acad. Sci. USA {\bf 112}, 3866 (2015).
	
	
	\bibitem{kubo2010}Y. Kubo, F. R. Ong, P. Bertet, D. Vion, V. Jacques, D. Zheng, A. Dr$\rm \acute{e}$au, J.-F. Roch, A. Auffeves, F. Jelezko, J. Wrachtrup, M. F. Barthe, P. Bergonzo, and D. Esteve, Strong Coupling of A Spin Ensemble to A Superconducting Resonator, Phys. Rev. Lett. {\bf105}, 140502 (2010).
	
	\bibitem{marcos2010}D. Marcos, M. Wubs, J. M. Taylor, R. Aguado, M. D. Lukin,and A. S. S\o rensen, Coupling Nitrogen-Vacancy Centers in Diamond to Superconducting Flux Qubits, Phys. Rev. Lett. {\bf105}, 210501 (2010).
	
	\bibitem{zhu2011} X. Zhu, S. Saito, A. Kemp, K. Kakuyanagi, S. Karimoto, H. Nakano, W. J. Munro, Y. Tokura, M. S. Everitt, K. Nemoto,
	M. Kasu, N. Mizuochi, and K. Semba, Coherent coupling of a superconducting flux qubit to an electron spin ensemble in diamond
	Nature (London) {\bf478}, 221 (2011).
	
	\bibitem{Qiu-2014}Y. Qiu, W. Xiong, L. Tian, and J. Q. You, Coupling spin ensembles via superconducting flux qubits, Phys. Rev. A {\bf89}, 042321 (2014).
	
	\bibitem{Lu-2013}X. Y. L$\rm \ddot{u}$, Z. L. Xiang, W. Cui, J. Q. You, and F. Nori, Quantum memory using a hybrid circuit with flux qubits and nitrogen-vacancy centers,  Phys. Rev. A {\bf88}, 012329 (2013).
	
	\bibitem{Mei-2009}F. Mei, M. Feng, Y. F. Yu, and Z. M. Zhang, Scalable quantum information processing with atomic ensemble and flying photons,  Phys. Rev. A {\bf80}, 042319 (2009).
	
	\bibitem{Mei-2010}F. Mei, Y. F. Yu, X. L. Feng, Z. M. Zhang, and C. H. Oh, Quantum entanglement distribution with hybrid parity gate,  Phys. Rev. A {\bf82}, 052315 (2010).
	
	\bibitem{Raizen-89} M. G. Raizen, R. J. Thompson, R. J. Brecha, H. J.
	Kimble, H. J. Carmichael, Normal-Mode Splitting and Linewidth Averaging for Two-State Atoms in An Optical Cavity, Phys. Rev. Lett. \textbf{63}, 240 (1989).
	
	\bibitem{Petersson-12} K. D. Petersson, L. W. McFaul, M. D. Schroer, M.
	Jung, J. M. Taylor, A. A. Houck, and J. R Petta, Circuit quantum electrodynamics with a spin qubit, Nature (London) \textbf{490}, 380 (2012).
	
	\bibitem{Rose-17}B. C. Rose, A. M. Tyryshkin, H. Riemann, N. V. Abrosimov, P. Becker, H.-J. Pohl, M. L. W. Thewalt, K. M. Itoh, and S. A. Lyon, Coherent Rabi dynamics of a superradiant spin ensemble in a microwave cavity, Phys. Rev. X {\textbf 7}, 031002 (2017).
	
	\bibitem{Dobrovitski-08} V. V. Dobrovitski, A. E. Feiguin, D. D. Awschalom,
	and R. Hanson, Decoherence dynamics of a single spin versus spin ensemble, Phys. Rev. B \textbf{77}, 245212 (2008).
	
	\bibitem{Wesenberg-09} J. H. Wesenberg, A. Ardavan, G. A. D. Briggs, J. J.
	L. Morton, R. J. Schoelkopf, D. I. Schuster, and K. M\o lmer, Quantum Computing with An Electron Spin Ensemble, Phys. Rev. Lett. \textbf{103}, 070502 (2009).
	
	\bibitem{Julsgaard-13}B. Julsgaard, C. Grezes, P. Bertet, and K. M\o lmer, Quantum Memory for Microwave Photons in An Inhomogeneously Broadened Spin Ensemble, Phys. Rev. Lett. \textbf {110}, 250503 (2013).
	
	\bibitem{Arcizet-2011}O. Arcizet, V. Jacques, A. Siria, P. Poncharal, P. Vincent, and S. Seidelin, A single nitrogen-vacancy defect coupled to a nanomechanical oscillator, Nat. Phys. {\bf7}, 879 (2011).
	
	\bibitem{Bennett-2013}S. D. Bennett, N.Y. Yao, J. Otterbach, P. Zoller, P. Rabl, and M. D. Lukin, Phonon-Induced Spin-Spin Interactions in Diamond Nanostructures: Application to Spin Squeezing, Phys. Rev. Lett. {\bf110}, 156402 (2013).
	
	\bibitem{Chotorlishvili-2013}L. Chotorlishvili, D. Sander, A. Sukhov, V. Dugaev, V. R. Vieira, A. Komnik, and J. Berakdar, Entanglement between nitrogen vacancy spins in diamond controlled by a nanomechanical resonator, Phys. Rev. B {\bf88}, 085201 (2013).
	
	\bibitem{Li-2016}P. B. Li, Z. L. Xiang, P. Rabl, and F. Nori, Hybrid Quantum Device with Nitrogen-Vacancy Centers in Diamond Coupled to Carbon Nanotubes, Phys. Rev. Lett. {\bf117}, 015502 (2016).
	
	\bibitem{Rabl-2009}P. Rabl, P. Cappellaro, M. V. Gurudev Dutt, L. Jiang, J. R. Maze, and M. D. Lukin, Strong Magnetic coupling between an electronic spin qubit and a mechanical resonator, Phys. Rev. B {\bf79}, 041302(R) (2009).
	
	\bibitem{Rabl-2010}P. Rabl, S. J. Kolkowitz, F. H. L. Koppens, J. G. E. Harris, P. Zoller, and M. D. Lukin, A quantum spin transducer based on nanoelectromechanical resonator arrays, Nat. Phys. {\bf6}, 602 (2010).
	
	\bibitem{Kolkowitz-2012}S. Kolkowitz, A. C. Bleszynski Jayich, Q. P. Unterreithmeier, S. D. Bennett, P. Rabl, J. G. E. Harris, and M. D. Lukin, Coherent sensing of a mechanical resonator with a single-spin qubit, Science {\bf335}, 1603 (2012).
	
	\bibitem{Pigeau-2015}B. Pigeau, S. Rohr, L. Mercier de Lepinay, A. Gloppe, V. Jacques, and O. Arcizet, Observation of a phononic mollow triplet in a multimode hybrid spin-nanomechanical system, Nat. Commun. {\bf6}, 8603 (2015).
	
	\bibitem{Li-2015}P. B. Li, Y. C. Liu, S. Y. Gao, Z. L. Xiang, P. Rabl, Y. F. Xiao, and F. L. Li, Hybrid quantum device based on NV centers in diamond nanomechanical resonators plus superconducting waveguide cavities, Phys. Rev. Applied {\bf4}, 044003 (2015).
	
    \bibitem{Zhou-2010}L. G. Zhou, L. F. Wei, M. Gao, and X. B. Wang, Strong coupling between two distant electronic spins via a nanomechanical resonator, Phys. Rev. A {\bf81}, 042323 (2010).

	\bibitem{MacQuarrie-2013} E. R. MacQuarrie, T. A. Gosavi, N. R. Jungwirth, S. A. Bhave, and G. D. Fuchs, Mechanical Spin Control of Nitrogen-Vacancy Centers in Diamond, Phys. Rev. Lett. {\bf111}, 227602 (2013).
	
	\bibitem{Teissier-2014}J. Teissier, A. Barfuss, P. Appel, E. Neu, and P. Maletinsky, Strain Coupling of A Nitrogen Vacancy Center Spin to A Diamond Mechanical Oscillator, Phys. Rev. Lett. {\bf113}, 020503 (2014).
	
	\bibitem{Ovartchaiyapong-2014}P. Ovartchaiyapong, K. W. Lee, B. A. Myers, and A. C. Bleszynski Jayich, Dynamic strain-mediated coupling of a single diamond spin to a mechanical resonator, Nat. Commun. {\bf5}, 4429 (2014).
	
	\bibitem{Kepesidis-2013}K. V. Kepesidis, S. D. Bennett, S. Portolan, M. D. Lukin, and P. Rabl, Phonon cooling and lasing with nitrogenvacancy centers in diamond, Phys. Rev. B {\bf88}, 064105 (2013).
	
	\bibitem{Lu-2015}X. Y. L$\rm \ddot{u}$, Y. Wu, J. R. Johansson, H. Jing, J. Zhang, and F. Nori, Squeezed Optomechanics with Phase-Matched Amplification and Dissipation, Phys. Rev. Lett. {\bf114}, 093602 (2015).
	
	\bibitem{Qin-2018} W. Qin, A. Miranowicz, P. B. Li, X. Y. L$\rm \ddot{u}$, J. Q. You, and F. Nori, Exponentially Enhanced Light-Matter Interaction, Cooperativities, and Steady-State Entanglement Using Parametric Amplification, Phys. Rev. Lett. {\bf120}, 093601 (2018).
	
	\bibitem{Xiong-2018}W. Xiong, Y. Qiu, L. A. Wu, and J. Q. You, Amplification of the coupling strength in a hybrid quantum system, New J. Phys. {\bf20}, 043037 (2018).
	
		\bibitem{Humphreys-2018}P. C. Humphreys, N. Kalb, J. P. J. Morits, R. N. Schouten, R. F. L. Vermeulen, D. J. Twitchen, M. Markham, and R. Hanson, Deterministic delivery of remote entanglement on a quantum network, Nature {\bf 558}, 268 (2018).
	
	
	\bibitem{Gao-2015}W. B. Gao, A. Imamoglu, H. Bernien, and R. Hanson, Coherent manipulation, measurement and entanglement of individual solid-state spins using optical fields, Nat. Photon. {\bf 9}, 363 (2015).
	
	
	\bibitem{Twamley2010}J. Twamley and S. D. Barrett, Superconducting cavity bus for single nitrogen-vacancy defect centers in diamond, Phys. Rev. B {\bf81}, 241202(R) (2010).
	
	\bibitem{Kippenberg-2008}T. J. Kippenberg and K. J. Vahala, Cavity optomechanics: back-action at the mesoscale. Science {\bf321}, 1172 (2008).
	
	\bibitem{Marquardt-2009}F. Marquardt and S. M. Girvin, Trend: Optomechanics. Physics {\bf2}, 40 (2009).
	
	\bibitem{Aspelmeyer-2012}M. Aspelmeyer, P. Meystre, and K. Schwab, Quantum optomechanics, Phys. Today {\bf65}, 29 (2012).
	
	\bibitem{Aspelmeyer-2014}M. Aspelmeyer, T. J. Kippenberg, and F. Marquardt, Cavity optomechanics, Rev. Mod. Phys. {\bf86}, 1391 (2014).
	
	\bibitem{Regal-2011}C. A. Regal and K. W. Lehnert,  From cavity electromechanics to cavity optomechanics. J. Phys. Conf. Ser. {\bf264}, 012025 (2011).
	
	\bibitem{Teufel1-2011}J. D. Teufel, D. Li, M. S. Allman, K. Cicak, A. J. Sirois, J. D. Whittaker, and R. W. Simmonds, Circuit cavity electromechanics in the strong-coupling regime. Nature {\bf471}, 204 (2011).
	
	\bibitem{Didier-2012}N. Didier and R. Fazio, Putting mechanics into circuit quantum electrodynamics. C. R. Phys. {\bf13}, 470 (2012).
	
	\bibitem{wangyindan}Y. D. Wang and A. A. Clerk, Using dark modes for high-fidelity optomechanical quantum state transfer, New J. Phys. {\bf14}, 105010 (2012).
	
	\bibitem{Vitali2007}D. Vitali, S. Gigan, A. Ferreira, H. R. B$\ddot{o}$hm, P. Tombesi, A. Guerreiro, V. Vedral, A. Zeilinger, and M. Aspelmeyer, Optomechanical Entanglement between a Movable Mirror and a Cavity Field, Phys. Rev. Lett. {\bf 98}, 030405 (2007).
	
	\bibitem{clerk2010}A. A. Clerk, M. H. Devoret, S. M. Girvin, Florian Marquardt, and R. J. Schoelkopf, Introduction to quantum noise, measurement, and amplification, Rev. Mod. Phys. {\bf 82}, 1155 (2010).
	
	\bibitem{Teufel2-2011} J. D. Teufel, T. Donner, D. Li, J. H. Harlow, M. S. Allman, K. Cicak, A. J. Sirois, J. D. Whittaker, K. W. Lehnert, and R. W. Simmonds, Sideband cooling of micromechanical motion to the quantum ground state. Nature {\bf475}, 359 (2011).
	
	\bibitem{Chan-2011} J. Chan, T. P. M. Alegre, A. H. Safavi-Naeini, J. T. Hill, A. Krause, S. Groeblacher, M. Aspelmeyer, and O. Painter, Laser cooling of a nanomechanical oscillator into its quantum ground state. Nature {\bf478}, 89 (2011).
	
	\bibitem{Weis-2010}S. Weis, R. Rivi$\grave{e}$re, S. Del$\acute{e}$glise, E. Gavartin, O. Arcizet, A. Schliesser, and T. J. Kippenberg, Optomechanically Induced Transparency, Science {\bf330}, 1520 (2010).
	
	\bibitem{Safavi-Naeini-2011}A. H. Safavi-Naeini, T. P. M. Alegre, J. Chan, M. Eichenfield, M. Winger, Q. Lin, J. T. Hill, D. E. Chang, and O. Painter, Electromagnetically induced transparency and slow light with optomechanics, Nature {\bf472}, 69 (2011).
	
	\bibitem{Brooks-2012}D. W. C. Brooks, T. Botter, S. Schreppler, T. P. Purdy, N. Brahms, and D. M. Stamper-Kurn, Non-classical light generated by quantum-noise-driven cavity optomechanics, Nature {\bf488}, 476 (2012).
	
	\bibitem{Safavi-Naeini-2013}A. H. Safavi-Naeini, S. Gr\"{o}blacher, J. T. Hill, J. Chan, M. Aspelmeyer, and O. Painter, Squeezed light from a silicon micromechanical resonator, Nature {\bf500}, 185 (2013).
	
	\bibitem{Bochmann-2013}J. Bochmann, A. Vainsencher, D. D. Awschalom, and A. N. Cleland, Nanomechanical coupling between microwave and optical photons, Nat. Phys. {\bf9}, 712 (2013).
	
	\bibitem{Andrews-2014}R. W. Andrews, R.W. Peterson, T. P. Purdy, K. Cicak, R.W. Simmonds, C. A. Regal, and K.W. Lehnert, Bidirectional and efficient conversion between microwave and optical light, Nat. Phys. {\bf10}, 321 (2014).
	
	\bibitem{Bagci-2014}T. Bagci, A. Simonsen, S. Schmid, L. G. Villanueva, E. Zeuthen, J. Appel, J. M. Taylor, A. S\o{}rensen, K. Usami, A. Schliesser, and E. S. Polzik, Optical detection of radio waves through a nanomechanical transducer, Nature (London) {\bf507},81 (2014).
	
	\bibitem{law1995} C. K. Law, Interaction between a moving mirror and radiation pressure: A Hamiltonian formulation, Phys. Rev. A {\bf51}, 2537 (1995).
	
	
	\bibitem{Niemczyk2009}T. Niemczyk, F. Deppe, M. Mariantoni, E. P. Menzel, E. Hoffmann, G. Wild, L. Eggenstein, A. Marx, and R. Gross, Fabrication technology of and symmetry breaking in superconducting quantum circuits, Supercond. Sci. Technol. {\bf22}, 034009 (2009).
	
	\bibitem{Niemczyk2010}T. Niemczyk, F. Deppe, H. Huebl, E. P. Menzel, F. Hocke, M. J. Schwarz, J. J. Garcia-Ripoll, D. Zueco, T. H$\rm \ddot{u}$mmer, E. Solano, A. Marx, and R. Gross, Circuit quantum electrodynamics in the ultrastrong-coupling regime, Nat. Phys. {\bf6}, 772 (2010).
	
	\bibitem{sm} See Supplementary Material at http://link.aps.org/
	supplemental/10.1103/PhysRevB.103.174106 for detailed
	derivations of our main results, which includes Refs. \cite{wangyindan,Vitali2007,clerk2010,Sudhir2012,lu2013sr,Frohlich,Nakajima,Angerer17}
	
	\bibitem{walls}D. F. Walls and G. J. Milburn, {\it Quantum optics} (Springer-Verlag, Berlin, 1994).
	
	\bibitem{zengzhao}Z. Z. Li, L. Bruder, F. Stienkemeier, and A. Eisfeld, Probing weak dipole-dipole interaction using phase-modulated nonlinear spectroscopy, Phys. Rev. A {\bf 95}, 052509 (2017).
	
	\bibitem{bin2018}Q. Bin, X. Y. L$\rm \ddot{u}$, L. L. Zheng, S. Wu. Bin, and Y. Wu, Detection of light-matter interaction in the weak-coupling regime by quantum light,Phys. Rev. A {\bf 97}, 043802 (2018).
	
	\bibitem{carreno2015}J. C. L. Carre$\tilde{n}$o, C. S. Mu$\tilde{n}$oz, D. S$\acute{a}$nchez Sanvitto, E. del Valle, and F. P. Laussy, Exciting Polaritons with Quantum Light, Phys. Rev. Lett. {\bf115}, 196402 (2015).
	
	\bibitem{Tekavec2007}P. F. Tekavec, G. A. Lott, and A. H. Marcus, Fluorescence-detected two-dimensional electronic coherence spectroscopy by acousto-optic phase modulation, J. Chem. Phys. {\bf127}, 214307 (2007).
	
	\bibitem{cuevas}$\acute{A}$. Cuevas, J. C. L. Carre$\tilde{n}$o, B. Silva, M. D. Giorgi,
	D. G. Su$\acute{a}$rez-Forero, C. S. Mu$\tilde{n}$oz, A. Fieramosca, F. Cardano,
	L. Marrucci, V. Tasco, G. Biasiol, E. del Valle, L. Dominici,
	D. Ballarini, G. Gigli, P. Mataloni, F. P. Laussy,
	F. Sciarrino, and D. S$\acute{a}$nchez Sanvitto, First observation of the quantized exciton-polariton field and effect of interactions on a single polariton, Sci. Adv. {\bf 4}, eaao6814 (2018).
	
	\bibitem{Sudhir2012}V. Sudhir, M. G. Genoni, J. Lee, and M. S. Kim, Critical behavior in ultrastrong-coupled oscillators, Phys. Rev. A {\bf 86}, 012316 (2012).
	
	\bibitem{lu2013sr}X. Y. L$\rm \ddot{u}$, W. M. Zhang, S. Ashhab, Y. Wu, and F. Nori, Quantum-criticality-induced strong Kerr nonlinearities in optomechanical systems, Sci.Rep. {\bf 3}, 2943,(2013).
	
	\bibitem{Angerer17} A. Angerer, S. Putz, D. O. Krimer, T. Astner, M. Zens, R. Glattauer, K. Streltsov, W. J. Munro, K. Nemoto, S. Rotter, J. Schmiedmayer, and J. Majer, Ultralong relaxation times in bistable hybrid quantum systems, Sci. Adv. {\bf 3}, e1701626 (2017).
	
	\bibitem{Frohlich}H. Fr\"{o}hlich, Theory of the Superconducting State. I. The Ground State at the Absolute Zero of Temperature, Phys. Rev. {\bf 79}, 845 (1950).
	
	\bibitem{Nakajima}S. Nakajima, Perturbation theory in statistical mechanics, Adv. Phys. {\bf 4}, 363 (1953).
	
	
	\bibitem{goppl2008}M. G$\rm \ddot{o}$ppl, A. Fragner, M. Baur, R. Bianchetti, S. Filipp, J. M. Fink, P. J. Leek, G. Puebla, L. Steffen, and A. Wallraff, Coplanar waveguide resonators for circuit quantum electrodynamics, J. Appl. Phys. {\bf104}, 113904 (2008).
	
	\bibitem{siampour}H. Siampour,  S. Kumar, and S. I. Bozhevolnyi, Chip-integrated plasmonic cavity-enhanced single nitrogen-vacancy center emission, Nanoscale {\bf9}, 17902 (2007).

	
	\bibitem{wang-2020}S. P. Wang, G. Q. Zhang, Y. Wang, Z. Chen, T. Li, J. S. Tsai, S. Y. Zhu, and J. Q. You, Photon-Dressed Bloch-Siegert Shift in an Ultrastrongly Coupled Circuit Quantum Electrodynamical System, Phys. Rev. Applied {\bf 13}, 054063 (2020).
	
	\bibitem{forn-2010} P. Forn-D\'iaz, J. Lisenfeld, D. Marcos, J. J. Garcia-Ripoll, E. Solano, C. J. P. M. Harmans, and J. E. Mooij, Observation of the Bloch-Siegert Shift in a Qubit-Oscillator System
	in the Ultrastrong Coupling Regime, Phys. Rev. Lett. {\bf 105},
	237001 (2010).
 \end{thebibliography}
\end{document}